\documentclass[ aps, showkeys, nofootinbib, floatfix, superscriptaddress]{revtex4}

\usepackage{amsfonts}

\usepackage{amssymb}
\usepackage{amsmath}
\usepackage{graphicx}
\begin{document}

\title{Linearized physics and gravitational-waves polarizations in the Palatini formalism of GBD theory}
 \author{Jianbo Lu}
 \email{lvjianbo819@163.com}
 \affiliation{Department of Physics, Liaoning Normal University, Dalian 116029, P. R. China}
  \author{Jiachun Li}
 \affiliation{Department of Physics, Liaoning Normal University, Dalian 116029, P. R. China}
  \author{Hui Guo}
 \affiliation{Department of Physics, Liaoning Normal University, Dalian 116029, P. R. China}
 \author{Zhitong Zhuang}
 \affiliation{Department of Physics, Liaoning Normal University, Dalian 116029, P. R. China}
\author{Xin Zhao}
 \affiliation{Department of Physics, Liaoning Normal University, Dalian 116029, P. R. China}

\begin{abstract}
 A generalized Brans-Dicke  (GBD)  theory in the framework of Palatini formalism are proposed in this paper. We derive the field equations by using the variational approach and obtain the linearized equations by using the weak-field approximation method.  We show various  properties of the geometrical scalar field in the Palatini-formalism of  GBD theory: it is massless and source-free, which are different from the results given in the metric-formalism of GBD theory. Also, we investigate the polarization modes of gravitational waves (GWs)  by using the the  geodesic deviation method and the Newman-Penrose method in the Palatini-GBD theory. It is observed  that there are three  polarizations modes  and four oscillation in the  Palatini-GBD theory. Concretely, they are the two transverse tensor ($+$) and ($\times$) standard polarization modes, and one breathing mode (with two oscillation).  The  results of GWs polarization in the Palatini-GBD theory are different from that in the metric-GBD theory, where there are four polarizations modes: the two standard tensorial modes ($+$ and $\times$), a scalar  breathing  mode, and  a massive scalar mode that is a mix of the longitudinal and the breathing polarization. Comparing with the Palatini-$f(\tilde{R})$ theory and the General Relativity, we can see that the  extra breathing mode   of GWs polarization   can be found in the Palatini-GBD theory.  At last, the expression of the parameterized post Newton (PPN) parameter is derived, which could pass through the experimental test.

\end{abstract}

%\pacs{98.80.-k}


\maketitle

\section{$\text{Introduction}$}

 General Relativity (GR) as the standard model of gravity theory is tested well, especially in the solar system.   But it also confronts some unanswered questions, such as the problems on  the dark matter \cite{DM1,DM2},  the late accelerating universe \cite{SN-acc1,SN-acc2}, the inflation \cite{inflation1,inflation2}, and the quantization, etc. Maybe, the gravity does not work in the framework of the GR theory at the large scales and the high-energy region. In order to explore the "ultimate" theory that adequately describe gravitational interaction, modified or extended gravity theories have been widely investigated \cite{mg1,mg2,mg3,mg4,mg5,mg6,mg7,mg8,mg9,mg10}, such as the $f(R)$ theory \cite{fr-review1,fr-review2,fr-3,fr-4}, the $f(T)$ theory \cite{ft1,ft2}, the $f(G)$ theory \cite{fg1,fg2}, the  Brans-Dicke (BD)  theory  \cite{original-BD}, and so on.

 In the BD theory, the Newton gravity constant $G(t)=1/\phi(t)$ is considered as a function of time. Thus, a scalar field $\phi$ can be introduced naturally into the action. Some observational constraints on $G(t)$  can be found in Refs. \cite{VG-MNRAS-2004-dwarf,VG-PRD-2004-white,VG-APJ,VG-PRD-2002-SN,VG-PRL-1996-neutron,vg-prl-constraint,vg-constraint-lu-prd,vg-constraint-lu-epjc}. The action of the BD theory has a form \cite{bd-action}
 \begin{equation}
 S_{BD}=\int \sqrt{-g}d^{4}x[\frac{1}{16\pi}(\phi R-\frac{\omega}{\phi}\partial_{\mu}\phi\partial^{\mu}\phi)+L_{m}],\label{action-BD}
 \end{equation}
 where $\omega$ is the coupling constant. For observational and theoretical  motivations, several extended versions of the BD theory have been developed, such as adding a potential term to the original BD theory \cite{BD-potential},   assuming the coupling constant $\omega$ to be variable with respect to time \cite{BD-omegat1,BD-omegat2}, generalizing $\phi$ to be a function $f(\phi)$ in the coupling term $\phi R$ \cite{BD-ST,BD-ST1}, choosing a higher-dimension geometry in the BD theory \cite{Ma-5DBD}, etc.  The applications of these extended BD theories have been investigated widely, such as at the aspects of cosmology \cite{GBD-cosmic1,GBD-cosmic2,GBD-cosmic3}, weak-field approximation \cite{GBD-weak}, observational constraints \cite{GBD-constraint1,GBD-constraint2}, and so on   \cite{BD-widely1,BD-widely2,BD-widely3}.

 Recently, a different strategy was proposed  to modify the BD theory  (called GBD)  in Refs. \cite{GBD-L,GBD}  by generalizing the Ricci scalar $R$ to be an arbitrary function $f(R)$ in the BD action.  Comparing with other modified theories,  one can find  that the GBD theory in the metric formalism have some interesting properties or solve some problems existing in other theories  \cite{GBD-L,GBD}. Several results could be exhibited  briefly as follows. (1)  The state parameter of geometrical dark energy in the GBD model can cross over the phantom boundary $w=-1$ as achieved in the quintom model,  without bearing the problems existing in the quintom model \cite{GBD} (in the double scalar-fields quintom model, it is required to include both the canonical quintessence field and the non-canonical phantom field in order to make the state parameter to cross over $w=-1$ \cite{quintom,quintom1}, while several fundamental problems are associated with phantom field, such as the problem of negative kinetic term and the fine-tuning problem, etc). (2) One knows that the metric-formalism $f(R)$ theories are equivalent to the BD theory with a potential (abbreviated as BDV) for taking a specific value of the coupling  parameter $\omega=0$ \cite{fr-review2}, where the specific choice: $\omega=0$ is quite exceptional, and it is hard to understand the corresponding absence of the kinetic-energy term for the field in the action. However,  the BD field in the GBD theory  owns the non-disappeared  kinetic term in the action \cite{GBD-L,GBD}. (3) Using the method of the weak-field approximation Ref. \cite{GBD-L} showed that the GBD theory could  solve the problem of $\gamma$ value emerging in $f(R)$ modified gravity (i.e. the inconsistent problem between the observational $\gamma$ value and the theoretical $\gamma$ value \cite{fr-review2,ppn-fr1,ppn-fr2}), without introducing the so-called chameleon mechanism. Here $\gamma$  is the parametrized post-Newtonian (PPN) parameter. Furthermore, the GBD theory tends to investigate the physics from the viewpoint of geometry, while the BDV or the two scalar-fields quintom model tends to solve physical problems from the viewpoint of matter. It is possible that several special characteristics of scalar field could be revealed through studies of the gravitational geometry in the GBD theory.

 It is well known that  some assumptions  have to be taken at prior for developing modified gravity theories of GR. For example: (1) which one or ones of the dynamical variables  (the metric, the connection,  or  the tetrad, etc.) should be chosen to describe the gravitational interaction? (2) As a covariant action theory, what is the constructed Lagrangian form for the gravitational system, and  which corresponding space-time geometry  (the Riemann, the Weitzenb$\ddot{o}$ck, the Riemann-Cartan, or the higher-dimension geometry, etc.) should be used? Then based on the constructed Lagrangian quantity,  one can derive the field equations by using the variational principle. And the surviving theories should pass through the tests of experiments from the solar system, the astrophysics, the cosmology, etc.

 As shown in some references, the  gravity theories defining in the Riemann geometry are often depicted in two formalisms: the metric formalism \cite{fr-review2,metric-variable} and the Palatini formalism \cite{Pala-variable,Pala-variable1}. In the metric formalism  (or the standard formalism) the Levi-Civta connection is related to metric, while in the Palatini formalism the metric and the connection are considered as the independent dynamical variables.  For a general non-linear $f(R)$ function, the different field equations can be obtained for the metric-$f(R)$ theories and the Palatini-$f(\tilde{R})$ theories \cite{fr-review2}, respectively. Comparing to the metric formalism, some  advantages of the Palatini approach could be found. For example: (1) the field equations  in the metric formalism are the fourth-order PDE (partial differential equation), while the field equations  in the Palatini formalism are the second-order PDE  which is easier to solve  and interpret \cite{Pala-merit1}; (2) Palatini-formalism theory raises the effective cutoff of the theory without introducing additional degrees of freedom below the Planck scale \cite{Pala-merit2};  (3)  Palatini-formalism theory could lead to different  inflationary predictions, which opens a door to test the nature of gravity by using the future cosmological observations \cite{Pala-merit3}; (4) Palatini-formalism theory could lead to different interactions among the Standard Model particles and the Higgs field  in the large field regime, with a potential impact on the entropy production process following the end of inflation \cite{Pala-merit4}, etc. \cite{Pala-merit5,Pala-merit6}.

 In this paper, we investigate the GBD theory in the framework of  Palatini formalism.  The field equations and the linearized equations are derived in the Palatini-GBD theory.   Comparing to the metric formalism of GBD theory  \cite{GBD-L}, some new results can be found in the Palatini-formalism of GBD theory. (1) The geometrical scalar field  in the Palatini-formalism of GBD theory is massless and source-free. These properties of geometric scalar field are different from the results given in the metric-formalism of GBD theory  \cite{GBD-L}; (2) The parameterized post Newton parameter in the Palatini-GBD theory has a form:  $\gamma=\frac{2\omega-F_{0}}{2\omega+F_{0}}$, which could pass through the experimental test: $ |\omega| > 40000$. Comparing to the Ref.  \cite{GBD-L}, we can see a different expression  of $\gamma$ in the metric-GBD theory, where $\gamma$  depends on the parameters: $\omega$, $F_{0}$ and $m_{s}$. (3) Given that investigating the polarization of  gravitational waves (GWs) can serve to discriminate the different gravitational theories, we also study the polarization modes of GWs in the Palatini-GBD theory. The  Newman-Penrose (NP) method \cite{gw-polar4,gw-polar5} and the geodesic deviation (GD) method  \cite{gw-polar6} are used to explore the polarization modes of GWs in the GBD theory.  It is observed  that  the polarization modes of GWs in the Palatini-GBD theory (the three  polarization types) are different from that in the metric-GBD theory (the four polarization types). The results in the GBD theory also show that   the extra scalar field $F$  and the BD scalar field generate the new polarizations  (the scalar polarization modes) for GWs  which are not present in the standard GR or the Palatini-$f(R)$ theory.

 The structure of our paper are as follows. In section II, we derive the basic equations in the  Palatini-GBD theory. In section III,  the linearized field equations are gained by using the method of the weak-field approximation.  Section IV  investigate the polarization modes of gravitational waves  in the Palatini-GBD theory.  Section V discuss the parameterized post Newton parameter (PPN). Section VI is the conclusions.

\section{$\text{Field equations  in the Palatini-formalism of GBD theory}$}

This section is devoted to derive the basic equations in the Palatini-GBD theory. The action of the GBD theory in the Palatini formalism read as,
\begin{equation}
S=S_g(g_{\mu \nu },\widetilde{\Gamma}^{\lambda}_{\mu\nu},\phi)+S_{\phi}(g_{\mu \nu },\phi)+S_m(g_{\mu \nu },\psi )=\frac{1}{2}\int d^4x{\cal L}_{T},\label{action}
\end{equation}
with the total Lagrange quantity
\begin{equation}
{\cal L}_T=\sqrt{-g}[\phi f(\tilde{R})- \frac{\omega}{\phi}\partial _\mu \phi \partial ^\mu \phi+\frac{16\pi }{c^4}{\cal L}_m].\label{lagrange}
\end{equation}
Here the metric $g_{\mu \nu }$ and the connection $\widetilde{\Gamma}^{\lambda}_{\mu\nu}$ are considered as the independent dynamical variables, $g$ denotes the determinant of $g_{\mu\nu}$, and ${\cal L}_m$ denotes the matter Lagrangian  associated with the matter field $\psi$ and $g_{\mu\nu}$. $f(\tilde{R})$ is an arbitrary  function of Ricci scalar: $\tilde{R}=g_{\mu\nu}\tilde{R}^{\mu\nu}$, and the Ricci tensor $\tilde{R}_{\mu\nu}$ is defined by the independent Palatini connection $\widetilde{\Gamma}^{\lambda}_{\mu\nu}$
\begin{equation}
\tilde{R}_{\mu\nu}=\tilde{R}^{\alpha}_{\mu\alpha\nu}=\partial_{\lambda}\widetilde{\Gamma}^{\lambda}_{\mu\nu}-\partial_{\mu}\widetilde{\Gamma}^{\lambda}_{\lambda\nu}
+\widetilde{\Gamma}^{\lambda}_{\mu\nu}\widetilde{\Gamma}^{\rho}_{\rho\lambda}-\widetilde{\Gamma}^{\lambda}_{\nu\rho}\widetilde{\Gamma}^{\rho}_{\mu\lambda}
\label{Ricci-Tensor-Pala}.
\end{equation}
 Using the variational principle,  we can derive the evolutional equations of the dynamical fields in the Palatini-formalism of GBD theory.  Varying the action (\ref{action}) with respect to $g_{\mu\nu}$ and $\phi$, we gain two field  equations   as follows
\begin{eqnarray}
\phi \left[ F(\tilde{R})\tilde{R}_{\mu \nu }-\frac{1}{2}f(\tilde{R})g_{\mu \nu }\right]+ \frac{1}{2}\frac{\omega}{\phi}g_{\mu\nu}\partial_\sigma\phi\partial^\sigma\phi
-\frac{\omega}{\phi}\partial_\mu\phi\partial_\nu\phi = 8\pi T_{\mu \nu },\label{gravitational-eq-Pala}
\end{eqnarray}
\begin{equation}
f(\tilde{R})+2\omega\frac{\Box \phi}{\phi} -\frac{\omega}{\phi^{2}}\partial _\mu \phi \partial ^\mu \phi=0,\label{BD-scalar-eq}
\end{equation}
where $F(\tilde{R})\equiv\partial f(\tilde{R})/\partial \tilde{R}$, $\Box \equiv \nabla ^\mu \nabla _\mu $ and  $T_{\mu \nu }=\frac{-2}{\sqrt{-g}}\frac{\delta S_m}{\delta g^{\mu \nu }}$ is the energy-momentum tensor of matter. The trace of Eq.(\ref{gravitational-eq-Pala}) is
\begin{eqnarray}
F(\tilde{R})\tilde{R}-2f(\tilde{R})+\frac{\omega}{\phi^{2}}\partial_\mu\phi\partial^\mu\phi = \frac{8\pi T}{\phi}.\label{trace-grav-Pala}
\end{eqnarray}
Varying the action with respect to $\widetilde{\Gamma}^{\lambda}_{\mu\nu}$ gives
\begin{eqnarray}
\widetilde{\nabla}_{\lambda}(\sqrt{-g}\phi F(\tilde{R})g^{\mu\nu})=0,\label{connection-eq}
\end{eqnarray}
where $\widetilde{\nabla}$ is the covariant derivative with respect to the Palatini connection.  Eq.(\ref{connection-eq})  implies that the connection can be represented as the Christoffel symbol associated with the metric $h_{\mu\nu}$ by defining $h_{\mu\nu}=\phi F(\tilde{R})g_{\mu\nu}$. Then we can  have a following relation
\begin{equation}
\widetilde{\Gamma}^{\lambda}_{\mu\nu}=\Gamma^{\lambda}_{\mu\nu}+\frac{1}{2\phi F}[-g_{\mu\nu}\partial^{\lambda}(\phi F)+\delta^{\lambda}_{\nu}\partial_{\mu}(\phi F)+\delta^{\lambda}_{\mu}\partial_{\nu}(\phi F)],\label{connection-relation}
\end{equation}
where $\Gamma^{\lambda}_{\mu\nu}$ is the Livi-Civita connection associated with the metric $g_{\mu\nu}$. Thus, by using Eq.(\ref{Ricci-Tensor-Pala}) the Ricci tensor and the Ricci scalar in the Palatini formalism are rewritten as
\begin{equation}
\tilde{R}_{\mu\nu}=R_{(g)\mu\nu}+\frac{3}{2(\phi F)^{2}}\nabla_{\mu}(\phi F)\nabla_{\nu}(\phi F)-\frac{1}{\phi F}\nabla_{\mu}\nabla_{\nu}(\phi F)-\frac{1}{2\phi F}g_{\mu\nu}\Box(\phi F),\label{Ricci-Tensor-metric}
\end{equation}
\begin{equation}
\tilde{R}=R_{(g)}+\frac{3}{2(\phi F)^{2}}\nabla^{\sigma}(\phi F)\nabla_{\sigma}(\phi F)-\frac{3}{\phi F}\Box(\phi F),\label{Ricci-scalar-metric}
\end{equation}
where $R_{(g)\mu\nu}$ and $R_{(g)}$ denotes the Ricci tensor and the  Ricci  scalar defining in the metric formalism, and all covariant derivatives are taken with respect to the metric $g_{\mu\nu}$. Combining above equations, the modified Einstein equation is derived as
\begin{equation}
G_{\mu\nu}=R_{(g)\mu\nu}-\frac{1}{2}R_{(g)}g_{\mu\nu}=\frac{8\pi T_{\mu\nu}}{\phi F}+8\pi T_{\mu\nu}^{eff}\label{Einstain-Tensor-Pala}
\end{equation}
with
\begin{equation}
\begin{split}
8\pi T_{\mu\nu}^{eff}=-\frac{\omega}{2\phi^2F}g_{\mu\nu}\partial_{\sigma}\phi\partial^{\sigma}\phi+\frac{f}{2F}g_{\mu\nu}+\frac{\omega}{\phi^{2} F}\partial_{\mu}\phi\partial_{\nu}\phi-\frac{3}{2(\phi F)^{2}}\nabla_{\mu}(\phi F)\nabla_{\nu}(\phi F)   \\
+\frac{1}{\phi F}\nabla_{\mu}\nabla_{\nu}(\phi F)-\frac{1}{2}g_{\mu\nu}\tilde{R}+\frac{3}{4(\phi F)^{2}}g_{\mu\nu}\nabla^{\sigma}(\phi F)\nabla_{\sigma}(\phi F)-\frac{1}{\phi F}g_{\mu\nu}\Box(\phi F).\label{re-gravitation-eq}
\end{split}
\end{equation}
When $f(\tilde{R})$ linear in $\tilde{R}$, Eq.(\ref{Einstain-Tensor-Pala}) is identical to the field equation in the metric formalism of the GBD theory \cite{GBD}. The trace of the gravitational field equation and the BD field equation have the forms:
\begin{equation}
\Box (\phi F)=\frac{8\pi T}{3}-\frac{\omega}{3\phi}\partial_{\mu}\phi\partial^{\mu}\phi+\frac{2\phi f(\tilde{R})}{3}-\frac{2\phi F}{3}\tilde{R}+\frac{1}{2\phi F}\nabla_{\mu}(\phi F)\nabla^{\mu}(\phi F)+\frac{\phi F}{3}R_{(g)},\label{trace-gravity-eq-re}
\end{equation}
\begin{equation}
\Box\phi-\frac{\partial_{\mu}\phi\partial^{\mu}\phi}{4\phi}=\frac{1}{4\omega}[8\pi T-\phi F R_{(g)}-\frac{3}{2\phi F}\nabla_{\mu}(\phi F)\nabla^{\mu}(\phi F)+3\Box (\phi F)].\label{re-BD-equation}
\end{equation}
 We can see that Eqs. (\ref{trace-gravity-eq-re}) and (\ref{re-BD-equation}) describe the dynamics of  the two scalar fields: $\phi$ and $F$ in the Palatini-GBD theory, which are different from the results in the  Palatini-$f(R)$ theory, where the scalar field carries no dynamics of its own  \cite{fr-review2}.

In this section, we derived the field equations of the  GBD theory by using the non-standard Palatini approach, where the connection was treated as an independent dynamical variable. The gravitational field equations in  this theory were gained by performing variations of action with respect to the metric and the connection, respectively. Variation with respect to the metric gave  new  field equation containing $F(\tilde{R})$, and variation with respect to the connection gave  the Riemann connection associated with metric $h_{\mu\nu}$ via appropriate conformal transformation. Based on the above field equations, we derive the linearized equations in the Palatini-GBD theory in the following.

\section{$\text{ Linearized field equations in the Palatini-GBD theory}$}

 Modified  gravitational  theory should have the correct weak-field limit at the Newtonian and the post-Newtonian levels.  In this section, we derive the linearized field equations in the Palatini-GBD theory via the weak-field  approximation method. And then in the following two sections, we solve the linearized field equations for two cases: the vacuum case and the static point-mass case, respectively.

 As a begining, we discuss the weak-field approximations of GBD theory in the Palatini formalism  via
\begin{eqnarray}
g_{\mu\nu}=\eta_{\mu\nu}+b_{\mu\nu},~~~~~~\phi=\phi_{0}+\varphi,~~~~~~F=F_{0}+\delta F,\label{weak-conditions}
\end{eqnarray}
where $\eta_{\mu\nu}$ denotes the Minkowski metric,  $\phi$  and $F$  are two scalar fields, and the following three relations are required: $|b_{\mu\nu}|\ll 1$,  $|\varphi|\ll \phi_{0}$ and $|\delta F|\ll F_{0}$. Using Eqs.(\ref{Einstain-Tensor-Pala}-\ref{weak-conditions}),   linearized field equations in the Palatini-GBD theory are derived as
 \begin{eqnarray}
\bar{R}_{(g)\mu\nu}-\frac{\bar{R}_{(g)}}{2}\eta_{\mu\nu}=
\partial_{\mu}\partial_{\nu}\frac{\delta F}{F_{0}}+\partial_{\mu}\partial_{\nu}\frac{\varphi}{\phi_{0}}
-\eta_{\mu\nu}\bar{\Box}_{p}\frac{\delta F}{F_{0}}-\eta_{\mu\nu}\bar{\Box}_{p}\frac{\varphi}{\phi_{0}}+\frac{8\pi T_{\mu\nu}}{\phi_{0}F_{0}},\label{eq-weak-gravity}
\end{eqnarray}
 \begin{eqnarray}
\bar{\Box}_{p}\varphi=\frac{3}{4\omega-3F_{0}}[8\pi T-\phi F_{0}\bar{R}_{(g)}+3\phi_{0}\bar{\Box}_{p}\delta F],\label{eq-weak-phi}
\end{eqnarray}
 \begin{eqnarray}
\bar{\Box}_{p}\frac{\delta F}{F_{0}}=\frac{8\pi T}{3\phi_{0}F_{0}}-\frac{\bar{\Box}_{p}\varphi}{\varphi_{0}}+\frac{\bar{R}_{(g)}}{3},\label{eq-weak-Phi}
\end{eqnarray}
where $\bar{\Box}_{p}=\partial^{\mu}\partial_{\mu}$. $\bar{R}_{(g)\mu\nu}$ and $\bar{R}_{(g)}$  denote  the linearized quantities, and they can be rewritten as
\begin{eqnarray}
\bar{R}_{(g)\mu\nu}=\frac{1}{2}(-2\partial_{\mu}\partial\nu b_{f}+2\partial\mu\partial\nu\frac{\varphi}{\phi_{0}}-\bar{\Box}_{p}\theta_{\mu\nu}+\frac{\eta_{\mu\nu}}{2}\bar{\Box}_{p}\theta-\eta_{\mu\nu}\bar{\Box}_{p} b_{f}+\eta_{\mu\nu}\bar{\Box}_{p}\frac{\varphi}{\phi_{0}}),\label{linear-Rmunu}
\end{eqnarray}
\begin{eqnarray}
\bar{R}_{(g)}=-3\bar{\Box}_{p} b_{f}+3\bar{\Box}_{p}\frac{\varphi}{\phi_{0}}+\frac{\bar{\Box}_{p}\theta}{2}\label{linear-R}
\end{eqnarray}
via introducing a new tensor $\theta_{\mu\nu}=b_{\mu\nu}-\frac{1}{2}\eta_{\mu\nu}b-\eta_{\mu\nu}\frac{\varphi}{\phi_{0}}+\eta_{\mu\nu}b_{f}$, where $b_{f}\equiv \frac{\delta F}{F_{0}}$, $b=\eta^{\mu\nu}b_{\mu\nu}$ and   $\theta=\eta^{\mu\nu}\theta_{\mu\nu}$.
After choosing a so-called Lorenz gauge or the Harmonic gauge: $\partial^{\nu}\theta_{\mu\nu}=0$ and using Eqs. (\ref{eq-weak-gravity}-\ref{linear-R}),  we get the linearized gravitational field equation and the linearized scalar-field equations in the Palatini-GBD theory as follows
\begin{eqnarray}
\bar{\Box}_{p}\theta_{\mu\nu}=-\frac{16\pi T_{\mu\nu}}{\phi_{0}F_{0}},\label{eq-box-theta}
\end{eqnarray}
\begin{eqnarray}
\bar{\Box}_{p}\varphi=\frac{4\pi T}{\omega},\label{eq-box-varphi}
\end{eqnarray}
\begin{eqnarray}
\bar{\Box}_{p} b_{f}=0,\label{eq-box-Phi}
\end{eqnarray}
with $T=\eta^{\mu\nu}T_{\mu\nu}$. Comparing the linearized field equation (\ref{eq-box-Phi}) in the Palatini-GBD theory with that in the metric-GBD theory: $\bar{\Box}_{p} b_{f}-m_{s}^{2}b_{f}=\frac{16\pi\omega T}{3\phi_{0}F_{0}(2\omega+3F_{0})}$, we can read some different properties for the scalar field $b_{f}$. In the  Palatini-formalism of GBD theory,  the scalar field $b_{f}$  is massless, while  in the metric-formalism GBD the  scalar field is massive.  We also can see that the scalar field $b_{f}$ is source-free in the Palatini-formalism GBD, which is different from the result in the metric-formalism GBD.

\section{$\text{Gravitational waves polarization  in the Palatini-GBD theory}$}

  Gravitational-waves (GWs) physics is an important aspect for probing the viable gravitational theory.   Studying on the polarization modes of GWs is also useful for exploring the valuable information on the early universe \cite{GW-polar}. How many additional polarization modes are detected   in  GWs experiments could instruct us to study which theories of gravity.  Given that more accurate observational data on  GWs  will be received in the future \cite{GW-observations-future}, it is  worthwhile to investigate GWs physics in alternative theories of gravity,  especially in the Palatini-formalism of modified gravity. The weak-field approximation method provides a natural way to study the GWs. And in some references, the authors have applied this method to discuss the polarization of GWs in different theories \cite{GW-other1,GW-other2,GW-other3,GW-other4,GW-other5,GW-other6,GW-other7,GW-other8}.

   Considering GWs which propagate along the $z$-direction, we have $k^{\alpha}=\varpi(1,0,0,1)$ with the angular frequency  $\varpi$. And let us consider an observer detecting the gravitational radiation described by a unit timelike vector: $u^{\alpha}=(1,0,0,0)$. In the vacuum, we solve the wave Eqs. (\ref{eq-box-theta}-\ref{eq-box-Phi}) in the Palatini-formalism GBD to get
\begin{equation}
\theta_{\mu\nu}=A_{\mu\nu}(\vec{p})\exp(i k_{\alpha}x^{\alpha}),\label{solution-theta}
\end{equation}
\begin{equation}
\varphi=c(\vec{p})\exp(i p_{\alpha}x^{\alpha}),\label{solution-varphi}
\end{equation}
\begin{equation}
b_{f}=d(\vec{p})\exp(i q_{\alpha}x^{\alpha}).\label{solution-hf}
\end{equation}
Where $k_{\alpha}$  denotes the four-wavevector, and it is a null vector with $\eta_{\mu\nu}k^{\mu}k^{\nu}=0$. Eq.(\ref{solution-theta}) denotes the plane-wave solution of gravitational radiation, while Eqs. (\ref{solution-varphi}) and (\ref{solution-hf}) denote the  plane-wave solutions for the massless BD-field perturbation $\varphi$ and the massless geometry-field perturbation $b_{f}$, respectively.

 In theory, several methods have been developed to analyze the polarization  of GWs \cite{gw-polar4,gw-polar5,gw-polar6,gw-polar1,gw-polar2,gw-polar3,GW-polar1a,GW-polar2a,GW-polar3a,GW-polar4a}, such as  the Newman-Penrose (NP) method \cite{gw-polar4,gw-polar5}, the  geodesic deviation (GD) method \cite{gw-polar1,gw-polar6}, etc. In the following,  we investigate  the polarization modes of GWs in the Palatini-GBD theory by using these two methods. In a local proper reference frame, the equation of geodesic deviation can be described as
 \begin{equation}
 \ddot{x}^{i}=-R^{i}_{~0k0}x^{k},\label{eq-GD}
 \end{equation}
 here $i$ and $k$ can be taken as $\{1,2,3\}$, respectively. $R^{i}_{~0k0}$  denotes the so-called "electric" components of the Riemann tensor  with its expression as follows \cite{GW-polar3a}
  \begin{equation}
 R^{(1)}_{i0j0}=(h_{i0,0j}+h_{0j,i0}-h_{ij,00}-h_{00,ij}),\label{linear-Riemann}
 \end{equation}
 where $h_{\mu\nu}$ denotes the linear perturbation. Using Eqs. (\ref{eq-GD}) and (\ref{linear-Riemann}), we gain
   \begin{eqnarray}
 \ddot{x}(t)=-(xh_{11,00}+yh_{12,00}),~~~~~~~~\nonumber\\
 \ddot{y}(t)=-(xh_{12,00}+yh_{11,00}),~~~~~~~~\nonumber\\
 \ddot{z}(t)=(2h_{03,03}-h_{33,00}-h_{00,33})z.\label{xyzdott}
 \end{eqnarray}
 Using solution (\ref{solution-theta}) and Eqs. (\ref{xyzdott}), we obtain
   \begin{eqnarray}
 \ddot{x}(t)=k_{0}^{2}[\hat{\epsilon}^{(+)}(k_{0})x+\hat{\epsilon}^{(\times)}(k_{0})y]\exp(ik_{\alpha}x^{\alpha})+c.c.,~~\nonumber\\
 \ddot{y}(t)=k_{0}^{2}[-\hat{\epsilon}^{(+)}(k_{0})y+\hat{\epsilon}^{(\times)}(k_{0})x]\exp(ik_{\alpha}x^{\alpha})+c.c.,\nonumber\\
 \ddot{z}(t)=0,~~~~~~~~~~~~~~~~~~~~~~~~~~~~~~~~~~~~~~~~~~~~~~~~~~~~~~~~\label{xyzdott-tensor}
 \end{eqnarray}
which describe the two standard plus and cross polarization modes of GR with the frequency $k_{0}$. For case of massless scalar field $\varphi$ we have
  \begin{eqnarray}
 \ddot{x}(t)=-p_{0}^{2}xc(\vec{p})\exp(ip_{\alpha}x^{\alpha})+c.c.,~~~~
 \ddot{y}(t)=-p_{0}^{2}yc(\vec{p})\exp(ip_{\alpha}x^{\alpha})+c.c.,~~~~
 \ddot{z}(t)=0,\label{xyzdott-bf}
 \end{eqnarray}
and for case of massless scalar field  $b_{f}$ we obtain
   \begin{eqnarray}
 \ddot{x}(t)=-q_{0}^{2}xd(\vec{p})\exp(iq_{\alpha}x^{\alpha})+c.c.,~~~~
 \ddot{y}(t)=-q_{0}^{2}yd(\vec{p})\exp(iq_{\alpha}x^{\alpha})+c.c.,~~~~
 \ddot{z}(t)=0.\label{xyzdott-varphi}
 \end{eqnarray}
Obviously, Eqs.(\ref{xyzdott-bf}) and (\ref{xyzdott-varphi}) indicate a breathing type of GWs polarization, which have two oscillation modes with the frequency $q_{0}$ and frequency $p_{0}$, respectively.  The same results can  also be obtained by using the Newman-Penrose (NP) method \cite{GW-polar3a,NP-tetrad}. Following the method shown in Refs. \cite{GW-polar3a,GW-polar4a,GNP-polar-six1,GNP-polar-six2,GNP-polar-six3,GNP-polar-six4,GNP-polar-six5,GNP-polar-six6,NP-polar-six}, the amplitudes of six polarizations in the Palatini-GBD theory can be calculated as follows
 \begin{eqnarray}
 p^{(l)}_{1}=-\frac{1}{6}R_{0303}=0,~~~~~~~~~~~~~~~~~~~~~~~~~~~~~~~~~~~~~~~~~~~~~~~~~~~~~~~~~~~~~~~~~~~~~\nonumber\\
 p^{(x)}_{2}=-\frac{1}{2}R_{0301}=0,~~~~~~~~~~~~~~~~~~~~~~~~~~~~~~~~~~~~~~~~~~~~~~~~~~~~~~~~~~~~~~~~~~~~~\nonumber\\
 p^{(y)}_{3}=\frac{1}{2}R_{0302}=0,~~~~~~~~~~~~~~~~~~~~~~~~~~~~~~~~~~~~~~~~~~~~~~~~~~~~~~~~~~~~~~~~~~~~~~~\nonumber\\
 p^{(+)}_{4}=-R_{0101}+R_{0202}=-2k_{0}^{2}\hat{\epsilon}^{(+)}(k_{0})\exp(ik_{\alpha}x^{\alpha})+c.c.,~~~~~~~~~~~~~~~~~~~~~\nonumber\\
 p^{(\times)}_{5}=2R_{0102}=2k_{0}^{2}\hat{\epsilon}^{(\times)}(k_{0})\exp(ik_{\alpha}x^{\alpha})+c.c.,~~~~~~~~~~~~~~~~~~~~~~~~~~~~~~~~~~\nonumber\\
 p^{(b)}_{6}=-R_{0101}-R_{0202}=2p_{0}^{2}c(\vec{p})\exp(ip_{\alpha}x^{\alpha})+2q_{0}^{2}d(\vec{p})\exp(iq_{\alpha}x^{\alpha})+c.c..
 \label{p-np}
 \end{eqnarray}
Here the six polarizations modes of GWs are: the longitudinal scalar mode $p_{1}^{(l)}$, the vector-$x$ model $p_{2}^{(x)}$, the vecotr-$y$ mode $p_{3}^{(y)}$, the plus tensorial  mode $p_{4}^{(+)}$, the cross  tensorial  mode $p_{5}^{(\times)}$, and the breathing scalar mode $p_{6}^{(b)}$, respectively. Form expressions (\ref{p-np}), we can read the  plus tensor polarization mode, the cross tensor polarization mode, and a breathing scalar mode with two oscillation in the Palatini-GBD theory.    It indicates that the extra scalar field $F$  and the BD scalar field generate the new polarizations of GWs which are not present in the standard GR or the Palatini-$f(\tilde{R})$ theory. In the GR and the Palatini-$f(\tilde{R})$ theory, both of them  predict two  tensorial polarization modes: $+$ and $\times$, not have any scalar modes \cite{GW-polar}.

 Comparing our results with other theoretical results in Refs. \cite{GW-polar,GW-polar1a,GW-polar2a,GW-polar3a,GW-polar4a}, it is observed that the polarization modes of GWs in the Palatini-GBD theory are  different from the results given in some other gravitational theories. For example, in the massive BD theory  \cite{GW-polar}, it has  two standard tensorial modes of GR and  two scalar modes (the longitudinal and the breathing modes); In the massless BD theory it owns two standard  tensorial modes and one breathing scalar mode \cite{GW-polar}; In the $f(R)$ theories, there are two tensorial modes and a massive scalar mode that is a mix of the longitudinal and the transverse breathing polarization \cite{GW-polar1a,GW-polar2a};  In the $f(T,B)$ theory of teleparallel gravity (it is equivalent to $f(R)$ gravity by linearized the field equations in the weak field limit approximation), there are three polarizations \cite{GW-polar3a}: the two standard of general relativity and an additional massive scalar mode, where the boundary term $B$ excites the extra scalar polarization; In the higher order local and non-local theories of gravity, they have three state of polarization and $n + 3$ oscillation modes \cite{GW-polar4a} (concretely, they are the two transverse tensor ($+$) and ($\times$) standard polarization modes of frequency $\omega_{1}$, and the $n+1$ further scalar modes of frequency $\omega_{2}, ..., \omega_{n+2}$, each of which has the same mixed polarization, partly longitudinal and partly transverse).

 We also compare the results of GWs  polarization in the Palatini-GBD theory with that in the metric-GBD theory. The plane-wave solutions in the  metric-GBD theory can be expressed as \cite{GBD-L}: $\theta_{\mu\nu}=A_{\mu\nu}(\vec{p})\exp(i k_{\alpha}x^{\alpha})$, $\varphi=a(\vec{p})\exp(i p_{\alpha}x^{\alpha})$, and $b_{f}=b(\vec{p})\exp(i q_{\alpha}x^{\alpha})$. Here,  $\varphi$ denotes the massless BD-field perturbation and $b_{f}$ denotes the massive geometry-field perturbation, respectively. For the massive plane wave propagating along $z-$direction, we have $q_{\alpha}=(q_{0},0,0,q_{3})$ with $ m^{2}=q_{0}^{2}-q_{3}^{2}\neq  0$. Originally, the NP formalism was applied to work out for massless waves. Recently, it was  also generalized  to explore the massive waves propagating along non-null geodesics \cite{GW-polar3a,NP-tetrad}. Using this method,  the non-zero amplitudes of polarizations for the metric-GBD theory in Ref. \cite{GBD-L} are calculated as
 \begin{eqnarray}
 p^{(l)}_{1}=\frac{1}{6}(-q_{3}^{2}+q_{0}^{2})b(\vec{p})exp(iq^{\alpha}x_{\alpha})+c.c.,~~~~~~~~~~~~~~~~~~\nonumber\\
 p^{(+)}_{4}=-\sqrt{2}k_{0}^{2}\hat{\epsilon}^{(+)}(k_{0})\exp(ik_{\alpha}x^{\alpha})+c.c.,~~~~~~~~~~~~~~~~~~\nonumber\\
 p^{(\times)}_{5}=\sqrt{2}k_{0}^{2}\hat{\epsilon}^{(\times)}(k_{0})\exp(ik_{\alpha}x^{\alpha})+c.c.,~~~~~~~~~~~~~~~~~~~~\nonumber\\
 p^{(b)}_{6}=2p_{0}^{2}a(\vec{p})\exp(ip_{\alpha}x^{\alpha})+2q_{0}^{2}b(\vec{p})\exp(iq_{\alpha}x^{\alpha})+c.c..
 \label{p-np-metric}
 \end{eqnarray}
 Obviously, Eqs.(\ref{p-np-metric})  show that there are four polarizations modes for  GWs in the metric-GBD theory: the two standard tensorial modes ($+$ and $\times$), a scalar  breathing  mode with frequency $p_{0}$, and  a massive scalar mode that is a mix of the longitudinal and the transverse breathing polarization with frequency $q_{0}$.

\section{$\text{ PPN parameter in the Palatini-GBD theory}$}

In this section, we  derive the theoretical expressions of the parametrized post-Newtonian (PPN) parameter $\gamma$  in the Palatini-GBD theory by using the weak-field approximation method.  Considering a static point-mass source, we have the form of the energy-momentum tensor: $T_{\mu\nu}=m_{p}\delta(\vec{r})diag(1,0,0,0)$. Obviously, the point particle is located at $\vec{r}=0$. Solving Eqs. (\ref{eq-box-theta}) and (\ref{eq-box-varphi}), we get the perturbation solutions: $\theta_{00}=\frac{4m_{p}}{\phi_{0}F_{0}}\frac{1}{r}$ and $\varphi(r)=\frac{m_{p}}{\omega}\frac{1}{r}$. Combining relations: $b_{\mu\nu}=\theta_{\mu\nu}-\eta_{\mu\nu}\frac{\theta}{2}+\eta_{\mu\nu}b_{f}-\eta_{\mu\nu}\frac{\varphi}{\phi_{0}}$ and $\theta=\eta^{\mu\nu}\theta_{\mu\nu}=-\frac{4m_{p}}{\phi_{0}F_{0}}\frac{1}{r}$, we gain  the non-vanishing components of the metric perturbation
\begin{equation}
b_{00}=\frac{2m_{p}}{\phi_{0}F_{0}r}+\frac{m_{p}}{\phi_{0}\omega r}-b_{f},\label{h00}
\end{equation}
\begin{equation}
b_{ij}=\frac{2m_{p}}{\phi_{0}F_{0}r}-\frac{m_{p}}{\phi_{0}\omega r}+b_{f}.\label{hij}
\end{equation}
Here $i,j =1,2,3$ is the space index. The term $\frac{m_{p}}{\phi_{0}\omega r}$ in above two equations reflect the effect  of scalar field $\phi$. Considering that $b_{f}$ is  negligible for a point-mass case, then the concrete form of the PPN  parameter $\gamma$  in the Palatini-GBD theory are derived as follows
\begin{equation}
\gamma=\frac{b_{ii}}{b_{00}}=\frac{2\omega-F_{0}}{2\omega+F_{0}}.\label{gamma}
\end{equation}
 From Eq. (\ref{gamma}), we can see the dependence of the PPN parameter $\gamma$ with respect to model parameters: $\omega$ and $F_{0}$. It is well known that a gravity theory alternative to GR should be tested by the well-founded experimental results. Some observations can be directly applied to constrain the value of the PPN parameter  $\gamma$. In Ref. \cite{bound-omega-gamma},  the experimental bound on $\gamma$ is: $|\gamma-1|<2.3*10^{-5}$. For the Palatini-GBD theory, then we have that $\gamma\sim 1$ requires $\omega \gg F_{0}$, which can be consistent with the observational constraint: $ |\omega| > 40000$ \cite{bound-omega-gamma}.

\section{$\text{Conclusions}$}

Several observational and theoretical  problems motivate us to investigate the  modified or alternative theories of GR.  Lots of modified gravity theories have been proposed and widely studied.  In the BD modified theory,  the scalar field can be introduced by considering a time-variable Newton gravity constant. Many extended versions of the BD theory have been explored and developed. As one of the generalized BD theories,  some interesting properties have been found in the metric-formalism of GBD theory \cite{GBD,GBD-L}. For examples: (1)  In the GBD model, the state parameter of geometrical dark energy can cross over the phantom boundary $w=-1$ as achieved in the quintom model,  without bearing the problems existing in the two-fields quintom model, such as the problem of negative kinetic term and the fine-tuning problem, etc. (2) It is well known that the  metric-$f(R)$ theories are equivalent to the BD theory with a potential (abbreviated as BDV) for taking a specific value of the BD parameter $\omega=0$, where the specific choice: $\omega=0$ for the BD parameter is quite exceptional, and it is hard to understand the corresponding absence of the kinetic term for the field. However, for the GBD theory, it is similar to the double scalar-fields model, and it owns the non-disappeared  kinetic term in the action.  In addition, the GBD theory tends to investigate the physics from the viewpoint of geometry, while the BDV or the two scalar-fields quintom model tends to solve physical problems from the viewpoint of matter. It is possible that several special characteristics of scalar field  could be revealed through studies of geometrical gravity in the GBD.

In this paper, we  studied the generalized Brans-Dicke theory in the Palatini formalism. Firstly, we derived the  field equations for  the gravitational field, the independent connection  and the BD scalar field, respectively. Secondly, using the weak-field method we obtained the  linearized gravitational field equation and the linearized scalar-fields equations. We showed various properties of the geometrical scalar field in the Palatini-formalism of GBD theory: it is massless and source-free, which are different from the results given in the metric-formalism of GBD theory.  According to the weak-field equations, we investigated the parameterized post Newton parameter in the Palatini-GBD theory by using the point-source method. It was shown that  $\gamma=\frac{2\omega-F_{0}}{2\omega+F_{0}}\sim 1$ requires $\omega \gg F_{0}$, which can be consistent with the observational constraint: $ |\omega| > 40000$.  Comparing to the Ref.  \cite{GBD-L}, we can see that the difference for expressions of $\gamma$ between the Palatini-GBD theory and the metric-GBD theory. In the metric-GBD theory  \cite{GBD-L},   $\gamma$  depends on the parameters: $\omega$, $F_{0}$ and $m_{s}$. Thirdly, we discussed the gravitational waves physics in the Palatini-GBD theory.  The properties of  GWs in the modified gravity theory have recently attracted lots of attention \cite{gw-mg1,gw-mg2,gw-mg3,gw-mg4,gw-mg5}, since investigating the polarization modes of GWs can serve to discriminate the different gravitational theories.
The results in the Palatini-formalism of GBD showed that   the extra scalar field $F$  and the BD scalar field generated the new polarizations for GWs which were not present in the standard GR or the Palatini-$f(R)$ theory, where both  theories (GR and Palatini-$f(\tilde{R})$) predicted two  tensorial polarization modes: $+$ and $\times$, not have any scalar modes. Concretely, we can read that there are three modes of polarization and four oscillation for GWs in the  Palatini-GBD theory, i.e. the plus tensor polarization mode, the cross tensor polarization mode, and a breathing scalar mode with two oscillation. The  results of GWs polarization in the Palatini-GBD theory are different from that in the metric-GBD theory. In the  metric-GBD theory, there are four polarizations modes: the two standard tensorial modes ($+$ and $\times$), a scalar  breathing  mode, and  a massive scalar mode that is a mix of the longitudinal and the breathing polarization.

\textbf{\ Acknowledgments }
 The research work is supported by   the National Natural Science Foundation of China (11645003,11705079), and supported by  LiaoNing Revitalization Talents Program.

\end{document}